\documentclass[prd,aps,preprint,tightenlines,showpacs,nofootinbib,superscriptaddress]{revtex4-1}
\usepackage{mathrsfs}
\usepackage{amsfonts}
\usepackage{amsmath}
\usepackage{amssymb}
\usepackage{array}
\usepackage{verbatim}
\usepackage{bm}
\usepackage{epsfig}
\usepackage{graphicx,color}
\usepackage{relsize}
\usepackage{lineno}
\usepackage{float}
\usepackage{multirow}
\RequirePackage{xspace}

\begin{document}


\title{\boldmath Investigating the Timelike Compton scattering in photon-proton interactions at the EIC}

\author{Ya-Ping Xie}
\email{xieyaping@impcas.ac.cn}
\affiliation{Institute of Modern Physics, Chinese Academy of Sciences,
	Lanzhou 730000, China}
\affiliation{University of Chinese Academy of Sciences, Beijing 100049, China}

\author{V.~P. Gon\c{c}alves}
\email{barros@ufpel.edu.br}
\affiliation{Institute of Physics and Mathematics, Federal University of Pelotas, \\
  Postal Code 354,  96010-900, Pelotas, RS, Brazil}
\affiliation{Institute of Modern Physics, Chinese Academy of Sciences,
  Lanzhou 730000, China}

\begin{abstract}
Information about the 3-dimensional description of the quark and gluon content of hadrons, described by the generalized parton distributions (GPDs), can be probed by exclusive processes in electron - proton ($ep$) collisions. In this letter, we investigate the timelike Compton scattering (TCS) in  $ep$ collisions at the future electron - ion collider (EIC). Such process is characterized by the exclusive dilepton production through the subprocess $\gamma p \rightarrow \gamma^*p \rightarrow l^+ l^- p$, with the real photon in the initial state being emitted by the incoming electron. Assuming a given model for the GPDs, the TCS differential cross-section is estimated, as well the contribution associated to the interference between the TCS and Bethe - Heitler (BH) amplitudes. Predictions for the TCS, BH and interference contributions are presented considering the kinematical range expected to be covered by the EIC detectors. Moreover, the polarized photon asymmetry is also studied. Our results indicated that a future experimental analysis, considering photon circular polarizations, can be useful to probe the interference contribution and constrain the description of the GPDs for the proton.
\end{abstract}

\maketitle

The  tomography picture from the hadrons can be revealed in  exclusive photon - hadron interactions, which are characterized by the fact that the hadron remains intact after scattering. Such processes can be studied in electron-hadron colliders as well as ultraperipheral collisions at RHIC and LHC (For recent reviews see, e.g. Refs. \cite{eic01, eic02, eic03,eic04,eic05,EicC,lhec01, lhec02}) and has a scattering amplitude that can be expressed in terms of  generalized parton distributions (GPDs) \cite{Muller:1994ses,Ji:1996ek,Radyushkin:1997ki,Collins:1998be}. As a consequence, they provide information about the 3-dimensional description of the quark and gluon content of hadrons \cite{Diehl:2003ny,Belitsky:2005qn}. Important constraints about the  GPDs have been derived from the  experimental results obtained in $ep$ collisions at HERA, JLab and COMPASS for the exclusive vector meson production, deeply virtual Compton scattering (DVCS) and for the timelike Compton scattering (TCS), with the promising expectation that future electron - ion colliders at BNL (EIC) \cite{eic04}, China (EicC) \cite{EicC}  and CERN (LHeC) \cite{lhec01, lhec02} will allow us to improve our understanding of the hadronic structure in a larger kinematical range.

\begin{figure}[!b]
	\centering
	\includegraphics[width=4.8in]{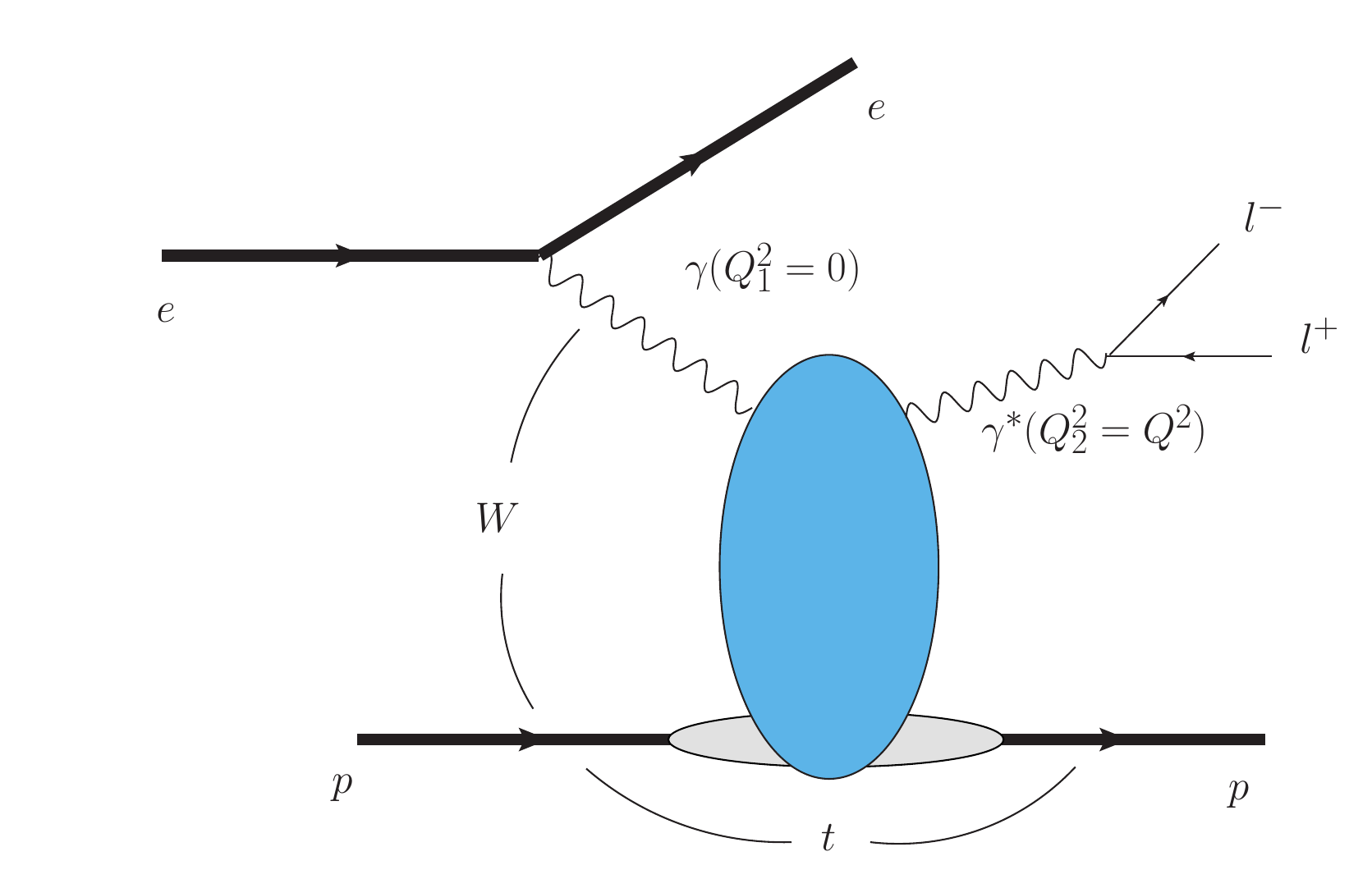}
	\caption{ Dilepton production in photon-proton interactions at the EIC.}
	\label{TCS}
\end{figure}

In this letter we will focus on the timelike Compton scattering (TCS), which is the exclusive photoproduction of a lepton pair with large invariant mass, considering the kinematical range that will be probed in $ep$ collisions at the EIC.  In previous investigations, TCS have been studied in LHC and JLab kinematics \cite{Pire:2008ea, Grocholski:2019pqj}.
Our goal is to complement the studies performed in Refs. \cite{Xie:2022vvl,Xie:2023vfi}, where we have demonstrated that the TCS process can be investigated in ultraperipheral $pA$ collisions at RHIC and LHC. In particular, we will consider the contribution induced by real photons, represented in Fig. \ref{TCS}, which is expected to be dominant in $ep$ collisions, and present predictions for the dependence of the differential cross-section on the squared transferred momentum $t$, azimuthal angle $\phi$, center of mass photon - proton energy $W$ and invariant mass of the dilepton system $Q^2$. Moreover, predictions for the photon beam circular polarization asymmetry will also be presented. 
We will show the TCS results, as well as those associated to the Bethe - Heitler (BH) process, which contributes to the same final state, and for the TCS - BH interference.

Initially, let's present a brief review of the formalism need to describe the TCS process in terms of generalized parton distributions (For a detailed discussion see, e.g., Refs. \cite{Diehl:2003ny,Belitsky:2005qn,Berger:2001xd,Mueller:2012sma,Boer:2015fwa}). As in these previous studies, we will assume the validity of the factorization theorem, which allow us to separate the hard scattering process, calculated using perturbation theory, from the non-perturbative dynamics encoded in GPDs. As in Ref. \cite{Xie:2022vvl}, in this exploratory study, the TCS process will be estimated at leading order of the strong running coupling constant $\alpha_s$, which implies that the Compton amplitude is dominated by the quark handbag diagrams described in terms of the quark GPDs\footnote{For a discussion of the next - to - leading order corrections and its impact see, e.g., Refs. \cite{Pire:2011st,Moutarde:2013qs,Xie:2023vfi}.}.
The impact of the contributions proportional to gluon GPDs will be presented in a forthcoming study.
  As the Bethe - Heitler (BH) process generates the same final state than the time-like Compton scattering, both processes contribute at the amplitude level, which implies that the differential cross-section for the dilepton production in  $ep$ collisions will be expressed as a sum of three contributions: $d\sigma^{total} = d\sigma^{TCS} + d\sigma^{BH} + d\sigma^{INT}$, where INT denotes the term associated to the TCS - BH interference. 
  The BH contribution had been estimated in Ref. \cite{Berger:2001xd}, where the differential cross-section is explicitly presented. In addition, as a detailed derivation of the main formulas associated to the TCS and INT contributions was presented in Refs. \cite{Berger:2001xd, Pire:2011st,Moutarde:2013qs}, here we will only present the final expressions, which have been used in phenomenological studies of the TCS process at the leading order in $\alpha_s$. 
  One has that the TCS contribution is given by\cite{Berger:2001xd}
\begin{eqnarray}
\frac{d\sigma^{TCS}_{\gamma \, p \rightarrow l^+l^- \,  p}}{dQ^{2}dtd(\cos\theta)d\phi} =  \frac{\alpha^3_{em}}{8\pi s_{\gamma p}^2}\frac{1}{Q^{2}}\frac{1+\cos^2\theta}{2}
\left\{(1-\eta^2)(|\mathcal{H}_1|^2 + |\widetilde{\mathcal{H}}_1|^2) \right. \nonumber \\
\left. -2\eta^2\mathrm{Re}[\mathcal{H}_1^*\widetilde{\mathcal{E}}_1]-\eta^2
\frac{t}{4M_p^2}|\widetilde{\mathcal{E}}_1|^2\right\}\,\,,
\label{Eq:TCS}
\end{eqnarray}
where $\eta = Q^2/(2s_{\gamma p} - Q^2)$,  $s_{\gamma p} = W^2$, $M_p$ is the proton mass and the Compton form factors 
$\mathcal{H}_1$, $\widetilde{\mathcal{H}}_1$,  $\mathcal{E}_1$, and $\widetilde{\mathcal{E}}_1$ are expressed in terms of the hard - scattering kernels $T_{\mathcal{H}_1,\widetilde{\mathcal{H}}_1,\mathcal{E}_1,\widetilde{\mathcal{E}}_1}^{q,g}$  and  the GPDs $H$, $\tilde{H}$, $E$ and $\tilde{E}$, defined in Ref. \cite{Diehl:2003ny} (For details, see Refs. \cite{Berger:2001xd,Pire:2008ea}). At leading order, one has that $T_{\mathcal{H}_1,\widetilde{\mathcal{H}}_1,\mathcal{E}_1,\widetilde{\mathcal{E}}_1}^{g} = 0$ and the expressions for the quark sector can be found in Refs. \cite{Berger:2001xd,Pire:2008ea}. The NLO kernels are presented in Refs.  \cite{Pire:2011st,Moutarde:2013qs}.
As statement in Ref. \cite{Berger:2001xd}, the $\theta$ in above equation is the angle between virtual photon and lepton in the lepton pair cms frame while the $\phi$ angle
is the angle between lepton pair plane and photon-proton plane. 
On the other hand, the contribution associated to the interference between the TCS and BH processes, considering  unpolarized protons and photons, can be expressed as follows \cite{Berger:2001xd,Lansberg:2015kha}
\begin{eqnarray}
\frac{d\sigma^{INT}_{\gamma \, p \rightarrow l^+l^- \,  p}}{dQ^{2}dtd(\cos\theta)d\phi}|_{\mbox{unpol}}  & = & -\frac{\alpha^3_{em}}{4\pi s^2_{\gamma p}}\frac{\sqrt{t_0 - t}}{-tQ}\frac{\sqrt{1 - \eta^2}}{\eta} \left(\cos\phi \frac{1+\cos^2\theta}{\sin\theta}\right) \nonumber \\
& \times & \mathrm{Re}\left[F_1(t)\mathcal{H}_1
-\eta(F_1(t) +F_2(t))\widetilde{\mathcal{H}}_1
-\frac{t}{4M_p^2}F_2(t)\mathcal{E}_1\right]\,\,,
\label{Eq:INT}
\end{eqnarray}
where $t_0 = - 4M_p^2 \eta^2/(1-\eta^2)$,  and we have neglected the lepton mass and assumed that 
$s_{\gamma p},\,Q^2 \gg t, M_p^2$. Moreover, $F_1(t)$ and $F_2(t)$ are the usual Dirac and Pauli form factors, with $F_2(0)$ normalized to the anomalous magnetic moment of the proton. One has that differently from the BH and TCS contributions, which are even under the transformation $\phi \rightarrow \phi + \pi$ for integration limits symmetric about $\theta = \pi/2$, the interference term is odd due to charge conjugation. Such aspect allow us to probe the Compton amplitude through a study of $\int_0^{2\pi} d\phi \cos \phi d\sigma/d\phi$.

One has that for the case of a scattering  between unpolarized protons and photons, one probes the real
part of the Compton helicity amplitudes.  On the other hand,  the
imaginary part can be accessed with polarized photon beams. If the photons have a circular polarization $\nu$, as is
the case for a bremsstrahlung beam emitted from longitudinally polarized leptons, one has that the interference term will be given by \cite{Berger:2001xd}
\begin{eqnarray}
\frac{d\sigma^{INT}_{\gamma \, p \rightarrow l^+l^- \,  p}}{dQ^{2}dtd(\cos\theta)d\phi}|_{\mbox{pol}}  & = & \frac{d\sigma^{INT}_{\gamma \, p \rightarrow l^+l^- \,  p}}{dQ^{2}dtd(\cos\theta)d\phi}|_{\mbox{unpol}} - \nu \frac{\alpha^3_{em}}{4\pi s^2_{\gamma p}}\frac{\sqrt{t_0 - t}}{-tQ}\frac{\sqrt{1 - \eta^2}}{\eta} \left(\sin\phi \frac{1+\cos^2\theta}{\sin\theta}\right) \nonumber \\
& \times & \mathrm{Im}\left[F_1(t)\mathcal{H}_1
-\eta(F_1(t) +F_2(t))\widetilde{\mathcal{H}}_1
-\frac{t}{4M_p^2}F_2(t)\mathcal{E}_1\right]\,\,.
\label{Eq:INT02}
\end{eqnarray}
One has that this interference contribution can be probed by investigating the photon beam circular polarization asymmetry, $A_{LU}$, defined by 
\begin{eqnarray}
A_{LU}= \frac{d\sigma(\nu =1)-d\sigma(\nu =-1)}{d\sigma(\nu = 1)+d\sigma(\nu =-1)}.
\end{eqnarray}

The main ingredient to estimate the TCS and INT cross-sections is a realistic model for the GPDs. In this exploratory study, we will consider the GPDs proposed and detailed in Refs. \cite{Goloskokov:2005sd,Goloskokov:2006hr,Kroll:2012sm}, usually called Goloskokov - Kroll (GK) model, which is based on fits of meson electroproduction data. In this model, the GPDs are expressed by
\begin{equation}
F_i(x, \xi, t) = \int_{-1}^1d\beta \int_{-1+|\beta|}^{1-|\beta|}d\alpha \, \delta(\beta + \xi \alpha -x)f_i(\beta, \alpha, t).
\end{equation}
where $F = H$, $\tilde{H}$, $E$, $\tilde{E}$ and $i$ denotes the quark flavor of the double distribution $f_i$, which are given by the following expressions
\begin{eqnarray}
f_i(\beta, \alpha, t) = g_i(\beta, t)h_i(\beta) \frac{\Gamma(2n_i+2)}{2^{2n_i+1}
	\Gamma^2(n_i+1)}\frac{[(1-|\beta|)^2-\alpha^2]^{n_i}}{(1-|\beta|)^{2n_i+1}}.
\end{eqnarray}
One has that $n_i$ is set to 1 for valence quarks and 2 for sea quarks. Moreover,  $h_{sea}^q (\beta,0) =q_{sea}(|\beta|)$sign$(\beta)$ and $h_{val}^q(\beta,0) = q_{val}(|\beta|)\Theta(\beta)$, 
where $q_{sea}$ and $q_{val}$ are the usual unpolarized PDFs. Finally, the $t$-dependence of the double distributions are described by $g_i(\beta, t)$, which is assumed to have a Regge behavior with linear trajectories, being given by
\begin{eqnarray}
g_i(\beta, t) = N e^{b_0 t}|\beta|^{-\alpha(t)}(1-\beta)^n.
\end{eqnarray}
The parameters considered in our analysis are detailed in Refs.\cite{Goloskokov:2005sd,Goloskokov:2006hr,Kroll:2012sm}. It is important to emphasize that the GK model provides a satisfactory description of the recent TCS data measured by the CLAS Collaboration \cite{CLAS:2021lky}.

\begin{figure}[t]
	\centering
	\includegraphics[width=3in]{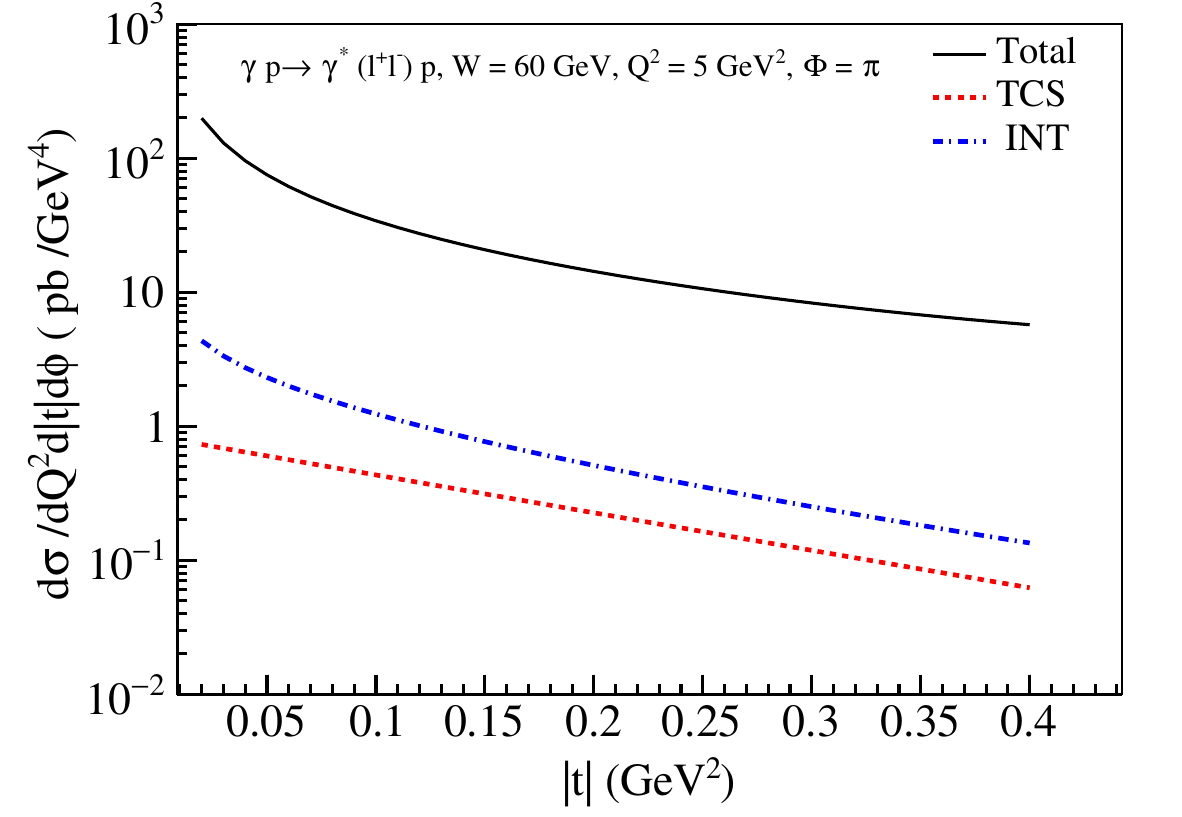}
	\includegraphics[width=3in]{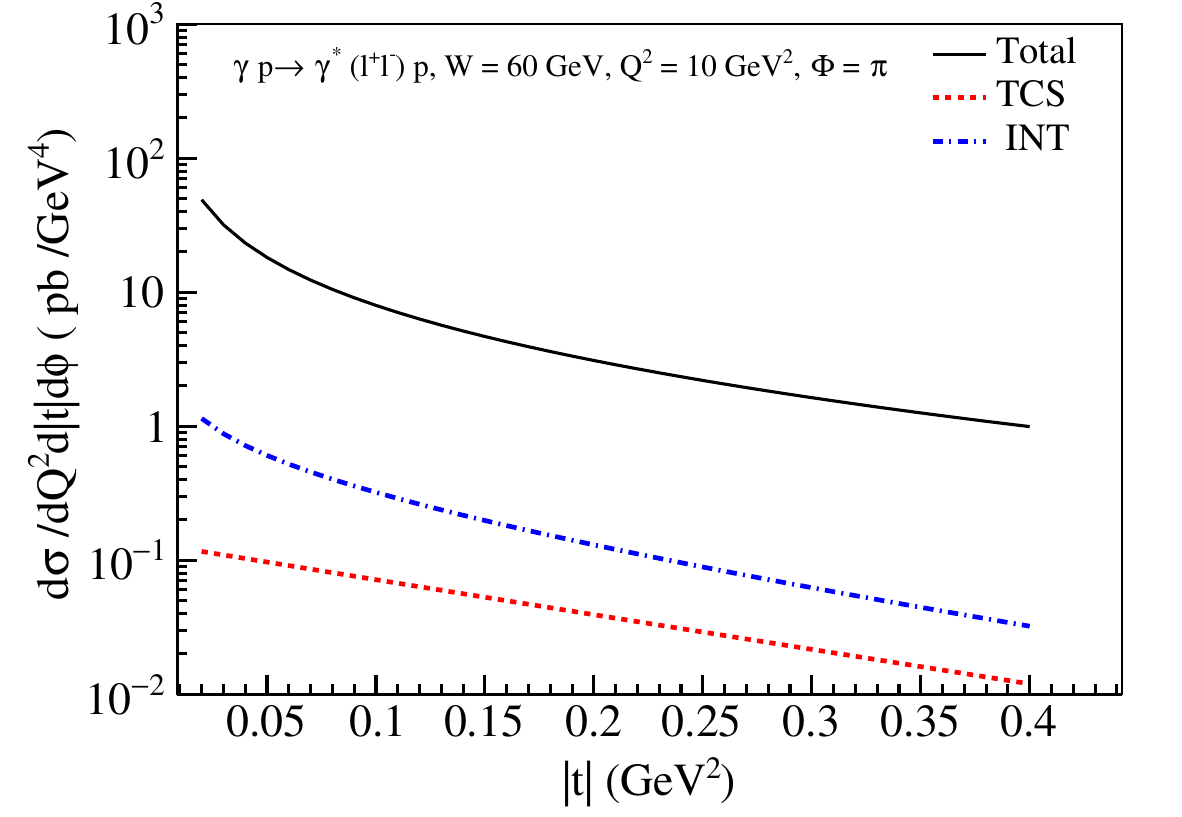}
	\caption{Predictions for $|t|$ dependence of the differential cross-section $d\sigma/dQ^2d|t|d\phi$ associated with the exclusive dilepton production in $ep$ collisions. The results for the total (BH + TCS + INT), TCS and INT contributions are presented separately. }
	\label{dsdt}
\end{figure}

In what follows, we will present our predictions for the exclusive dilepton production considering the TCS, BH and interference contributions. We will initially consider  unpolarized protons and photons and assume $W = 60$ GeV, which is a typical value for the photon - proton center - of - mass energy, expected to be probed in future $ep$ collisions at the EIC.
The results for the $|t|$ - distribution, obtained considering two values of $Q^2$, fixed azimuthal angle $\phi$ and  integrating $\theta$ from $\pi/4$ to $3\pi/4$, are presented in Fig.~\ref{dsdt}. In agreement with the predictions derived in previous studies \cite{Berger:2001xd,Pire:2008ea,Xie:2022vvl},  the BH contribution is larger than the TCS and INT contributions. However, our results indicate that the  INT contribution is not negligible in the kinematical range considered. Moreover, one has that the cross-section decreases with the increasing of $|t|$ and $Q^2$.

\begin{figure}[t]
	\centering
\includegraphics[width=3in]{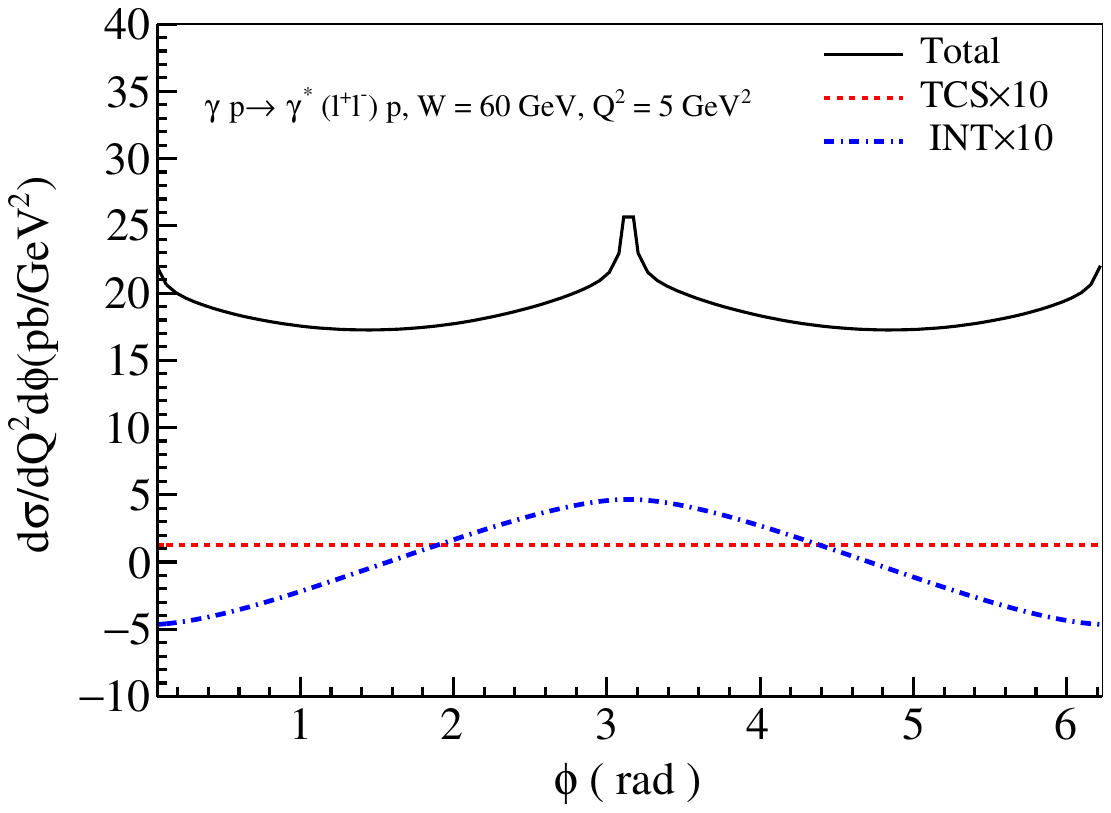}
\includegraphics[width=3in]{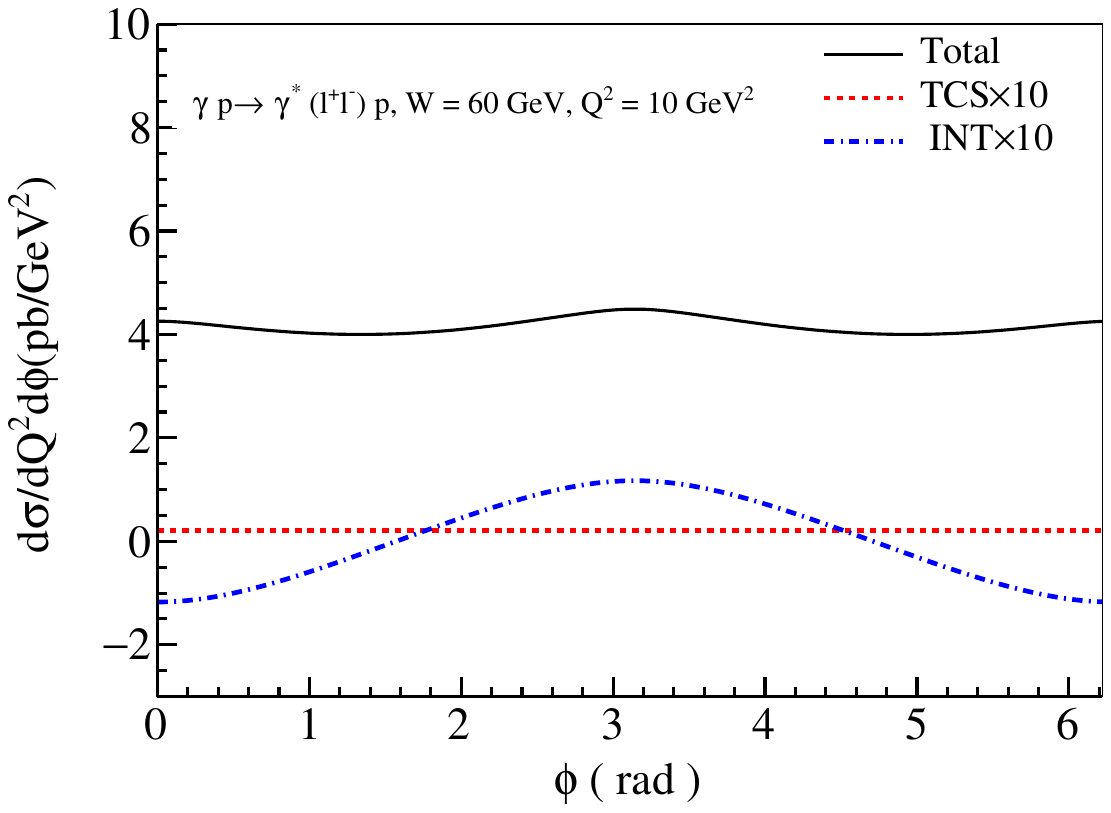}
\caption{Predictions for azimuthal angle distribution of the differential cross-section $d\sigma/dQ^2d\phi$ associated with the exclusive dilepton production in $ep$ collisions. The results for the total (BH + TCS + INT), TCS and INT contributions are presented separately.}
	\label{dsdphi}
\end{figure}

In Fig.~\ref{dsdphi} we present our predictions for the dependence on $\phi$ of the differential cross-section, derived considering $W$ = 60 GeV, Q$^2$ = 5 and 10 GeV$^2$, $\theta$ integrated from $\pi/4$ to $3\pi/4$ and $|t|$ over the full kinematical range. One has that TCS contribution  is a constant, since it is independent of $\phi$. On the other hand, as the unpolarized INT contribution is proportional $\cos\phi$, one has that it is larger for $\phi \approx \pi$. We have verified that the INT contribution is about 4 percent of the total cross-section in the kinematical range considered. Thus, this contribution is not negligible in the total cross-section. Moreover, one has that the peak for $\phi \approx \pi$ in larger for smaller values of the dilepton invariant mass $Q^2$.

\begin{figure}[t]
	\centering
	\includegraphics[width=3.2in]{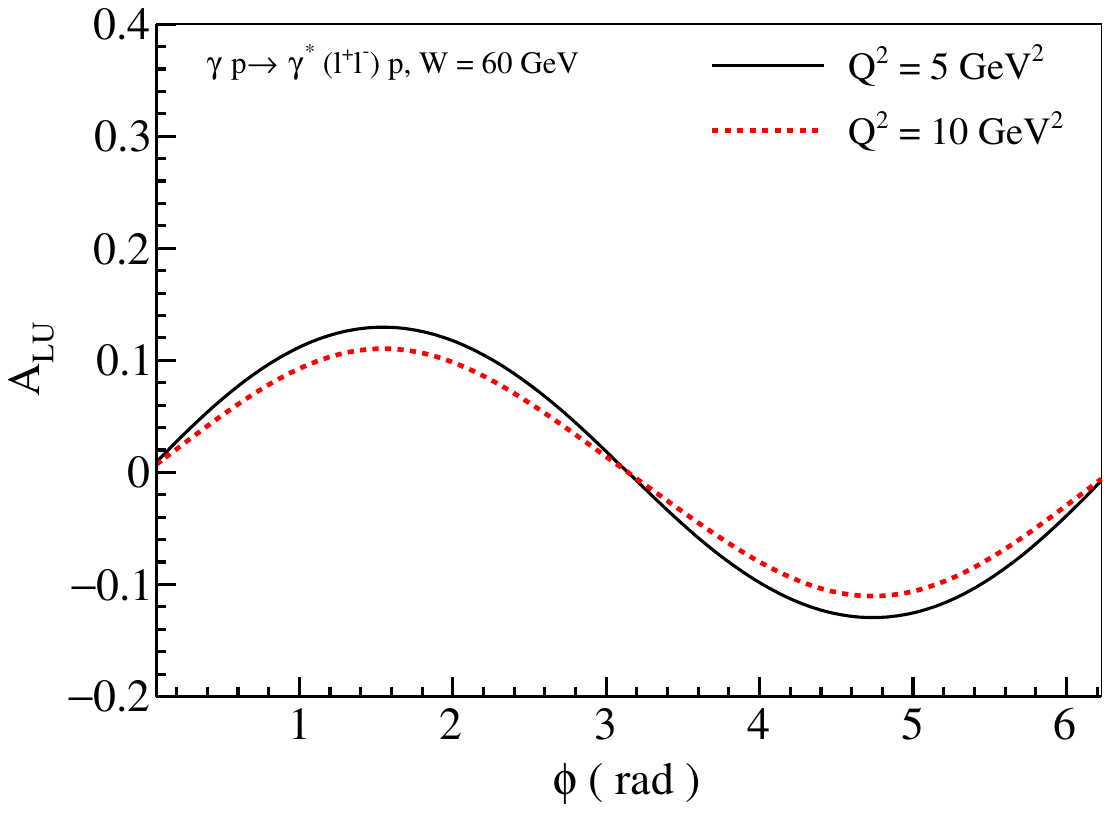}
	\includegraphics[width=3.2in]{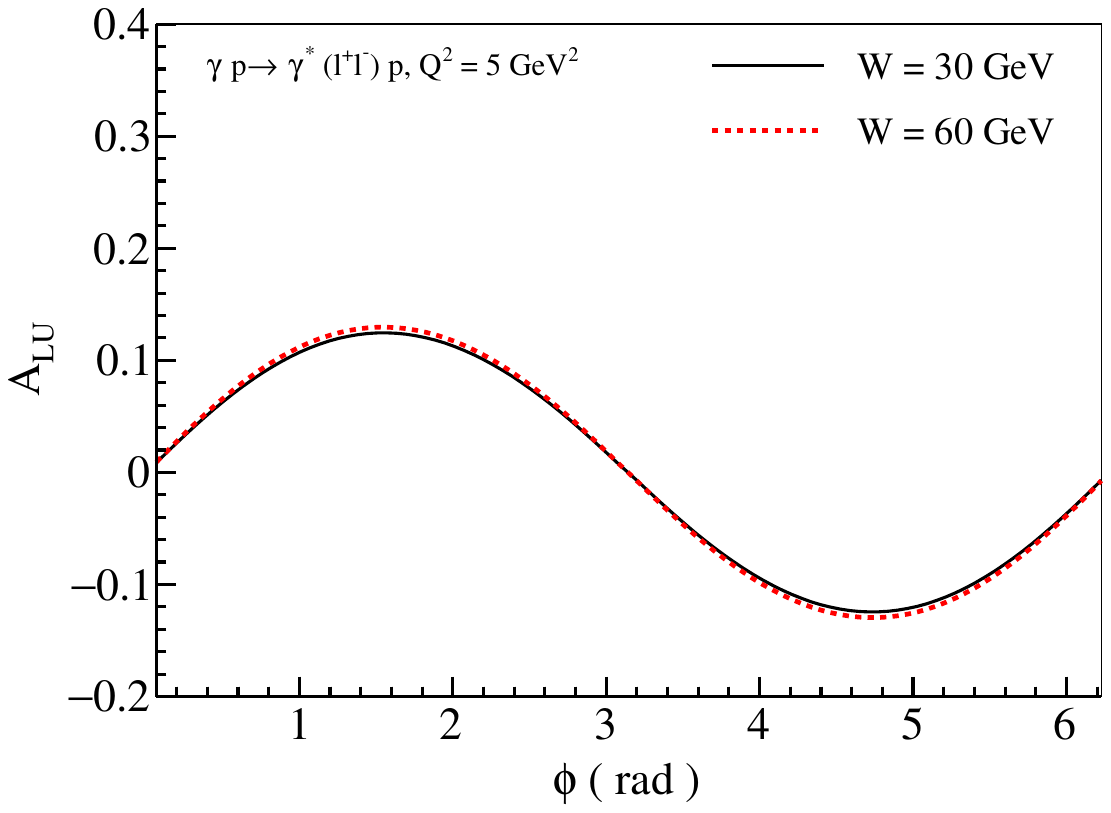}
	\caption{Predictions for the dependence of $A_{LU}$ on the azimuthal angle $\phi$ for different values of $Q^2$ (left panel) and distinct photon - proton center - of - mass energies (right panel). }
	\label{ALU01}
\end{figure}

Let's now analyze the behavior of the observable $A_{LU}$, which can accessed making use of photon circular polarizations. The  $\phi$ dependence of the photon polarized asymmetry $A_{LU} (W, Q^2, \phi)$ is presented in Fig.~\ref{ALU01}, derived by integrating $\theta$ in the range  [$\pi$/4, $3\pi$/4] and $|t|$ in the full kinematical range, for two values of $Q^2$ (left panel) and two values of $W$ (right panel). The proportionality of $A_{LU}$ with  $\sin\phi$ is clearly observed. One has that it reaches $\approx 10 \%$ for $Q^2$ = 5 GeV$^2$, decreasing for larger values of $Q^2$. On the other hand, one has that $A_{LU}$ is almost energy independent in the range considered.

\begin{figure}[t]
	\centering
	\includegraphics[width=3.2in]{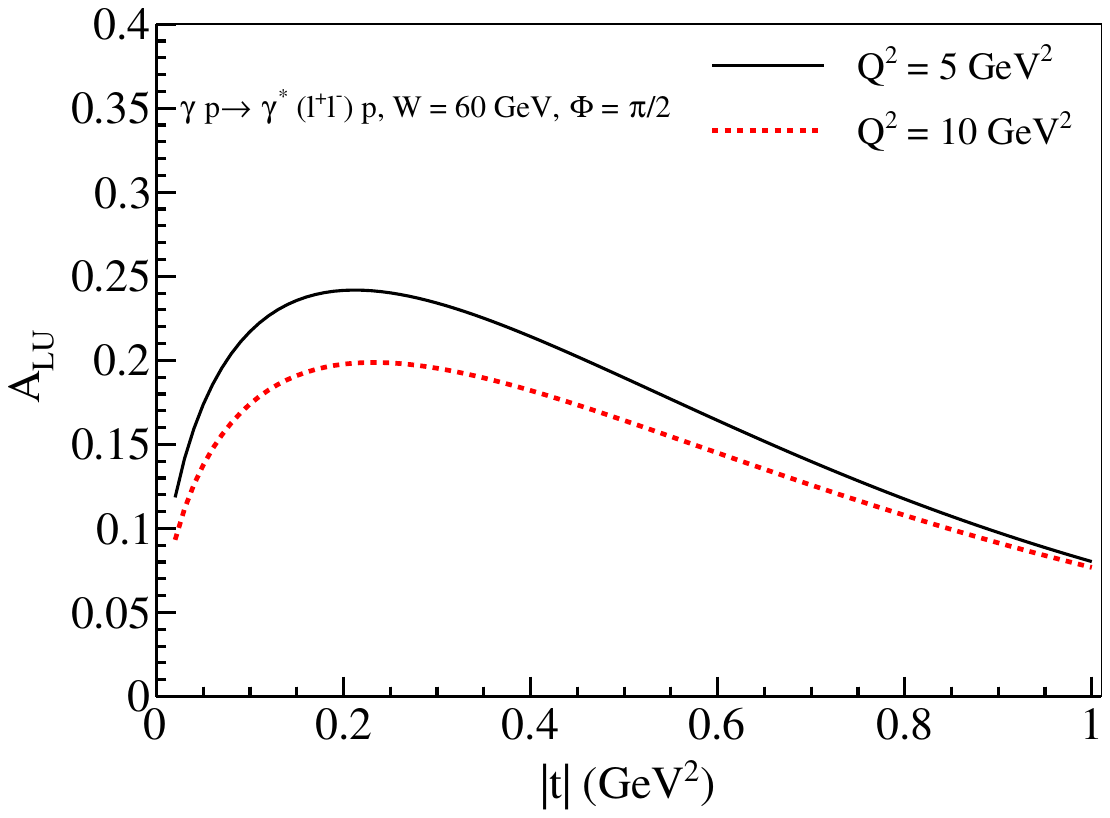}
	\includegraphics[width=3.2in]{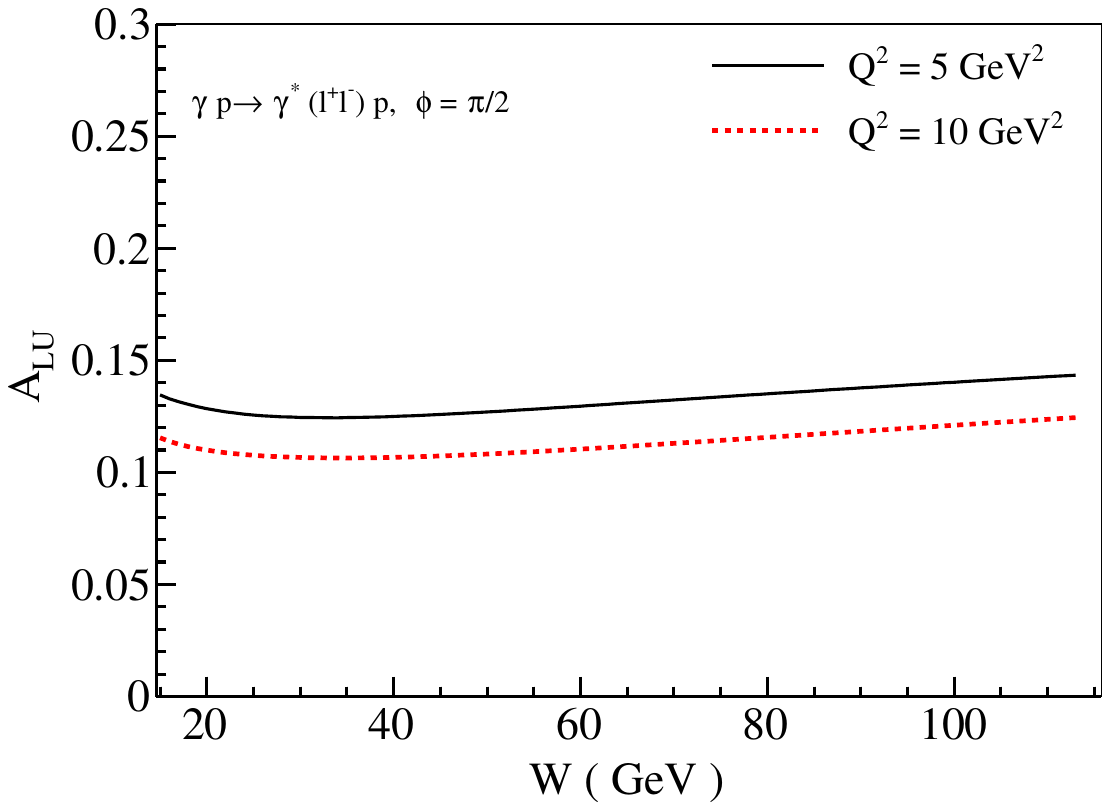}
	\caption{Predictions for the dependence of $A_{LU}$ on the squared transferred momentum (left panel) and $W$ (right panel) for two values of $Q^2$ and $\phi = \pi/2$. }
	\label{ALU02}
\end{figure}

Finally, in Fig.~\ref{ALU02}, we present our predictions for the $|t|$ (left panel) and $W$ (right panel) dependencies of   $A_{LU} (W, Q^2, |t|, \phi)$. One has that the largest value of $A_{LU}$ occurs for  $|t|\approx$ 0.2 GeV$^2$. Moreover, the dependence on $W$ is similar for the two values of $Q^2$ considered, with the normalization decreasing with $Q^2$.  These results indicate that a future measurement of the  photon beam circular polarization asymmetry
for small values of the invariant mass of the dilepton system and $|t|\approx$ 0.2 GeV$^2$ can be useful to probe the INT contribution and probe the description of the GPDs.

As a summary, the improvement of  our understanding about the quantum 3D imaging of the partons inside the protons and nuclei is one  of the main goals of Particle Physics. The quantum information of how partons are distributed inside hadrons is encoded in the
quantum phase space Wigner distributions, which include information on both generalized parton distributions (GPDs)
and transverse momentum parton distributions (TMDs). In this letter, we have investigated 
the timelike Compton scattering, which is one of the promising process to probe
 the quark and gluon GPDs of the proton. In particular, we focus on the kinematical range that will be probed by the future Electron - Ion Collider at the BNL/USA.  The contributions of the Bethe - Heitler, TCS and interference were estimated at leading order in $\alpha_s$ 
  and predictions for the kinematical range expected to be covered by the future EIC detectors were presented assuming the GK model for the quark GPDs.  Our results indicate that the 
  exclusive dilepton production in  $ep$ collisions is dominated by the BH contribution. However, if this contribution is subtracted, our results indicate that the TCS 
  and INT contributions are not negligible and can be probed in future experimental analysis.  We also provided predictions for the photon beam circular polarization asymmetry, $A_{LU}$, which indicated that it is larger for smaller values of $|t|$ and $Q^2$. Such results indicate that a future experimental determination of the INT contribution is, in principle, feasible.

\section*{Acknowledgments}
The work is partially supported by the NFSC grant (Grant No 12293061) and the Strategic Priority Research Program of Chinese Academy of Sciences (Grant No. XDB34030301).
 V.P.G. was partially supported by the CAS President's International Fellowship Initiative (Grant No.  2021VMA0019) and by CNPq, CAPES, FAPERGS and  INCT-FNA (process number 
464898/2014-5).

\end{document}